\documentclass{ws-rv9x6}
\pdfoutput=1
\usepackage{ws-rv-van}             
\makeindex
\begin{document}

\chapter{Hybrid quantum systems of atoms and ions\\ \label{chapter}}

\author{Carlo Sias} \address{Istituto Nazionale di Ottica -
CNR\\
via G. Sansone 1, IT-50019 Sesto Fiorentino-Firenze, Italy\\
carlo.sias@ino.it}

\author[Carlo Sias and Michael K\"ohl]{Michael K\"ohl}
\address{Physikalisches Institut, Universit\"at Bonn\\
Wegelerstrasse 8, 53115 Bonn, Germany\\
and\\
Cavendish Laboratory, University of Cambridge\\
JJ Thomson Av. Cambridge CB30HE, United Kingdom\\
michael.koehl@uni-bonn.de}
\begin{abstract}

In this chapter we review the progress in experiments with hybrid systems of trapped ions and ultracold neutral atoms. We give a theoretical overview over the atom-ion interactions in the cold regime and give a summary of the most important experimental results. We conclude with an overview of remaining open challenges and possible applications in hybrid quantum systems of ions and neutral atoms.

\end{abstract}

\body

\section{Introduction}\label{sec1.1SiasKohl}

In recent decades, scientists have learned how to experimentally create quantum systems and isolate them from the environment. These quantum systems can be controlled and manipulated at an extraordinary level, so that now experiments are possible which probe genuine quantum properties, such as quantum phases or quantum dynamics. With the excellent control at hand, one might also consider hybrid experiments in which distinct quantum system are coupled with each other. This is still a pioneering field, since the experiments are extremely demanding, and only a few hybrid quantum systems have been successfully realized so far.
Among them, hybrid quantum systems of ultracold atoms and trapped ions take a pivotal role since they combine the best-controlled many-body and single-body quantum systems.

What are the advantages of merging these two distinct fields and approaching them in the same setup? In order to answer this question, consider first the level of experimental control that has been reached in the two separate fields. On the one hand, with regard to ultracold atoms, other chapters in this book describe the impressive progress in the preparation of dilute gases of neutral atoms that has lead to engineering Hamiltonians by manipulating atom interactions, motion, dimensionality, etc. On the other hand, with regard to trapped ions, experimentalists are able to isolate a single particle, trap it efficiently for long periods of time (even a few months), cool it down to its motional ground state, and perform precise spectroscopic measurements on it. Trapped ions are, for instance, the building blocks of the best clock ever built\cite{wineland08}$.$ Additionally, genuine quantum many-body states like maximally entangled states have been realized in small crystals of cold trapped ions. The quantum computer with the largest number of qubits reported so far is based on trapped ions\cite{blatt11}$,$ and trapped ions are the first massive particles whose states have been successfully teleported\cite{blatt04,wineland04}$.$

Joined together, ions and neutral atoms can be conceived as probes and systems at the same time. From the point of view of the atoms, a single ion would act as a single, localized impurity in a many-body system. One can imagine the ion acting as a coherent probe in the gas, measuring local densities and correlations,
or the state of localized neutral atoms in an optical lattice\cite{kohl07}$.$ From the point of view of an ion, the atoms act as an ultracold
bath\cite{vuletic09,zipkes10}$,$ which is, in principle, transparent to the laser light used to manipulate it. Continuous cooling of trapped ions would increase the efficiency of quantum computation, since it would not be necessary to stop the computation in order to cool down the particles in an ongoing experiment. The ion could act as the reaction center for chemical processes in order to create molecules in a controlled way\cite{ratschbacher12,willitsch11}$.$ These molecules can in principle become large clusters acting as traps for the atoms, exploiting the atom-ion interaction potential\cite{cote02}$.$

The aim of this chapter is to present the basics of
atom-ion physics at cold temperatures, and to summarize the first key
experiments realized in hybrid systems of ultracold atoms and trapped ions. We will first discuss the basics of two-body atom-ion collisions, starting from the interaction potential and deriving the scattering cross section. Next, we will introduce the
basic concepts in conceiving a hybrid experimental setup in which both ions and atoms can be trapped and made to interact. Then, we will discuss the first pioneering experiments with ions and atoms, from sympathetic cooling of ions in a Bose-Einstein condensate\cite{zipkes10}$,$ to the controlled chemical interaction\cite{ratschbacher12} with the creation of molecular ions\cite{willitsch11}$,$ to the coherent evolution of an
ion spin qubit in an environment of ultracold atoms\cite{ratschbacher13}$.$ Finally, we will give an overview of the perspectives and open challenges that persist in this field.

\section{Basic theory of atom-ion interactions}\label{sectioninteractionSiasKohl}

In order to access the physics of atoms and ions at cold temperature it is essential to study the fundamental two-body atom-ion interaction. The leading energy scale results from the effects of the ion's charge onto the atom. An ion of charge $Q$ creates at a distance $R$ an electric field of amplitude $\left|\mathcal{E}_{ion}\right|=Q/\left(4\pi \epsilon_0 R^2\right)$, where $\epsilon_0$ is the vacuum permittivity. This electric field polarizes the neutral atom, creating an induced dipole of amplitude $\alpha_0 \left|\mathcal{E}_{ion}\right|$, where $\alpha_0$ is the atom's static electric polarizability. The electric field and the induced dipole interact so that the atom and the ion experience a potential \cite{massey34}
\begin{equation}
V\left(R\right)=-\frac{C_4}{2R^4},\label{R4PotentialSiasKohl}
\end{equation}
where $C_4=\alpha_0 Q^2/\left(4 \pi \epsilon_0\right)^2$. It is interesting to note that the characteristic length scale of the atom-ion potential is much more long-ranged than the Van-der-Waals interaction between two neutral atoms. For instance, the characteristic radius of the potential $R^*=\sqrt{\mu C_4/\hbar^2}$ is for Rb-Yb$^+$ interactions $R^*=307$nm. Here $\mu=\left(m_{ion}m_{at}\right)/\left(m_{ion}+m_{at}\right)$ is the reduced mass and $m_{ion}$ and $m_{at}$ are the masses of the ion and the neutral atom, respectively.

\begin{figure}
\centerline{\psfig{figure=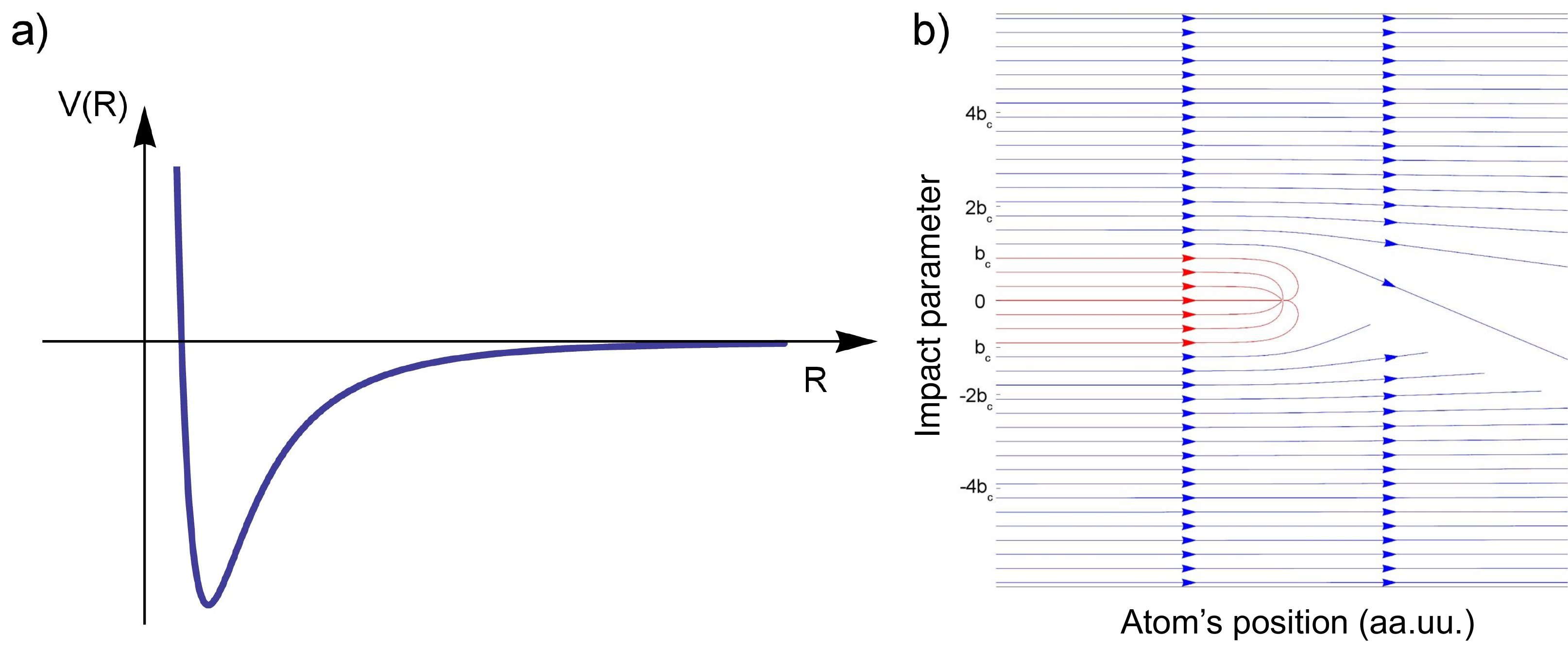,width=\textwidth}}
\caption{  a) Ion-atom interaction potential vs internuclear separation $R$.   b) Classical trajectories of an atom colliding with a trapped ion. When the impact   parameter is smaller than the critical value $b_c$ the atom's trajectory   spirals down towards the ion causing a collision   at a very short distance. }
\label{figinteractionSiasKohl}
\end{figure}

The potential in eq. (\ref{R4PotentialSiasKohl}) diverges for $R\rightarrow 0$; this unphysical effect is corrected by the presence of a hard-core short-range repulsion, as shown in Fig. \ref{figinteractionSiasKohl}(a). It is necessary to use quantum mechanics and to know exactly the shape of the hard-core repulsion in
order to calculate correctly the effects of the collisions. Nevertheless, it is sufficient to use classical physics and the simple model of equation (\ref{R4PotentialSiasKohl}) to observe the main features of atom-ion interactions\cite{langevin05}$.$ These are shown in Fig. \ref{figinteractionSiasKohl}(b), in which the different trajectories of an atom scattering with an ion at a  fixed collision energy $E_{coll}$ are presented for different impact  parameters\cite{wannier54} $b$, i.e. for different minumum distances between the unperturbed trajectories of the colliding particles. The special scaling $R^{-4}$ causes the presence of two distinguishable sets of trajectories that are experienced by the colliding particles depending whether $b$ is bigger or smaller than a critical impact parameter defined as: $b_c= \left(2 C_4/E_{coll}\right)^{1/4}$. If $b<b_c$, the trajectory of the colliding particles spirals down to short interparticle distances, and the outgoing wavepackets are directed almost isotropically in space. We will refer to these collisions as Langevin collisions\cite{langevin05}$.$ Contrarily, if $b\rangle b_c$, the free-particle trajectories are only slightly modified by the interaction. We will therefore refer to these collisions as ``forward scattering''. Since particles can exchange a considerable amount of momentum only in Langevin collisions, one can assume that Langevin collisions are the only ones contributing to the total cross section. The Langevin cross section reads:
\begin{equation}
\sigma_{Lang}=\pi b_c^2=\pi \left(\frac {2 C_4}{E_{coll}}\right)^{1/2}.\label{langcrosssectSiasKohl}
\end{equation}
It is important to note that the Langevin collision cross section scales with the inverse of the square root of the collision energy $E_{coll}^{-1/2}$.
This implies that the collisional rate constant $K_{Lang}=\sigma_{Lang} v$, where $v$ is the relative velocity of the colliding particles, is independent of the collision energy $E_{coll}$. Therefore, the collisional rate $\Gamma_{coll}=n K_{Lang}$ depends only on fundamental properties of the atom-ion system as $C_4$ and $\mu$, and on the particle density $n$. This is a special feature of the $R^{-4}$ interaction potential.

How is a quantum mechanical calculation different with respect to the classical picture of atom-ion collisions? In order to answer this question, it is convenient to calculate the collisional cross section by using an expansion in partial waves\cite{massey34,dalgarno00}
\begin{equation}
\sigma_{el}=\sum_{l=0}^{\infty}\sigma_{l}=\frac{4\pi}{k^2}\sum_{l=0}^{\infty} \left(2l+1\right)\sin^2\left(\eta_l\right),
\label{partialwaveexpansionSiasKohl}
\end{equation}
where $\hbar k=\sqrt{ 2\mu E_{coll}}$ is the momentum of the colliding particles in the center of mass frame, $\eta_l$ is the $l$-th partial wave phase shift, and the subscript in $\sigma_{el}$ indicates that we are considering elastic collisions in which the energy of the colliding particles is not transferred from/to different electronic energy levels. The phase shift for each partial wave can be calculated by solving the corresponding Schr\"odinger equation
\begin{equation}
\left(\frac{d}{dR^2}+k^2-\frac{2\mu}{\hbar^2}V_{int}\left(R\right)-\frac{l\left(l+1\right)}{R^2}\right)\psi\left(R\right)=0,
\end{equation}
where the last term in brackets is the centrifugal barrier for angular momentum $l \hbar$, and $V_{int}\left(R\right)$ is the full interaction potential including the potential (\ref{R4PotentialSiasKohl}) and the short-range repulsive hard wall. Without considering for the moment effects of tunneling of the centrifugal barrier, we can see that for a given collisional energy $E_{coll}$, there is a momentum $L=\left(1/\hbar\right)\sqrt{2\mu\sqrt{2 C_4 E_{coll}}}$ such that for $l>L$ the particles collide from the centrifugal barrier, while for $l<L$ the particles' energy is sufficiently high to pass over the centrifugal barrier so that the collision is from the hard-wall. These two cases correspond to the classical cases of forward scattering and Langevin collisions, respectively.

In order to determine the total elastic cross section, one has to compute separately the different contributions for $l>L$ and $l<L$ in eq. (\ref{partialwaveexpansionSiasKohl}). In case $l<L$, i.e. a Langevin collision, it is necessary to know exactly the expression for the hard wall potential in order to evaluate correctly the phase shifts $\eta_l$. Without this knowledge, one can assume that the phase shifts are isotropically distributed between $0$ and $2\pi$, so that one can assume $\sin^2\left(\eta_l\right)\simeq1/2$ in eq. (\ref{partialwaveexpansionSiasKohl}). With this assumption, which is better valid the bigger $L$ is, each partial wave contributes $\sigma_{l<L}=2\pi l/\left(k^2\right)$. Summing $\sigma_{l<L}$ for all $l$s up to $L$ gives the Langevin cross section of eq.(\ref{langcrosssectSiasKohl}). On the other hand, in order to calculate the contributions from partial waves with $l>L$ in eq. (\ref{partialwaveexpansionSiasKohl}), one can
consider the phase shifts in
the semiclassical approximation
\begin{equation}
\eta_{l>L}\simeq -\frac{\mu}{\hbar^2}\int_{R_0}^\infty \frac{V\left(R\right)}{\sqrt{k^2-\left(\frac{l+\frac{1}{2}}{R^2}\right)^2}}\simeq \frac{\pi \mu^2 C_4}{4 \hbar^4} \frac{E_{coll}}{l^3},\label{semiclassapproxSiasKohl}
\end{equation}
where $R_0=(l+1/2)/k$ is the classical turning point. If $L$ is sufficiently large, one can approximate $\left(2l+1\right)\sin^2\left(\eta_l\right)\sim2\eta_l^2$ in eq. (\ref{partialwaveexpansionSiasKohl}) for $l>L$. Using the formula of the
right hand side of 
eq.(\ref{semiclassapproxSiasKohl}), one
can evaluate the contribution to the total elastic cross section from forward scattering. i.e. for $l>L$. Summing the contributions for $l<L$ and $l>L$ in eq. (\ref{partialwaveexpansionSiasKohl}), we obtain the total elastic cross section, which reads
\begin{equation}
\sigma_{el}\left(E_{coll}\right)=\pi \left(\frac{\mu C_4^2}{\hbar^2}\right)^{1/3}\left(1+\frac{\pi^2}{16}\right)E_{coll}^{-1/3}.\label{totalelcrosssectSiasKohl}
\end{equation}
The total elastic cross section scales with $E_{coll}^{-1/3}$, so the elastic collision rate constant $K_{el}=\sigma_{el} v\propto E_{coll}^{1/6}$ depends on energy. This behavior is qualitatively different with respect to Langevin collisions, which are anyway included in the calculation of $\sigma_{el}$.

All these calculations are valid for collisional energies $E_{coll}$ sufficiently large so that several partial waves contribute to the cross section. This regime characterizes current experiments\cite{vuletic09}$,$ since the long-range feature of the interaction potential causes the s-wave scattering limit to occur at temperatures much lower than currently available\cite{idziaszek09}$.$ Moreover, for higher densities it is possible to observe effects whose explanation goes beyond the two-body atom-ion collision theory that we have treated here. A first evidence was the observation of an enhancement of three-body atom losses in a Rb-Rb$^+$ system\cite{denschlag10} and even more complicated many-body effects like the creation of mesoscopic molecular ions have been predicted\cite{cote02,gao10}$.$

\section{Building a hybrid system of atoms and ions}

In order to perform experiments with trapped ions and neutral atoms in the cold regime, it is necessary to build a hybrid setup in which atoms and ions can be cooled and trapped in the same physical location. To this end, two independent trapping potentials have to be created. These potentials can be magnetic or optical for the neutral atoms, and electrical or optical for the ions.

\begin{figure}
\centerline{\epsfig{figure=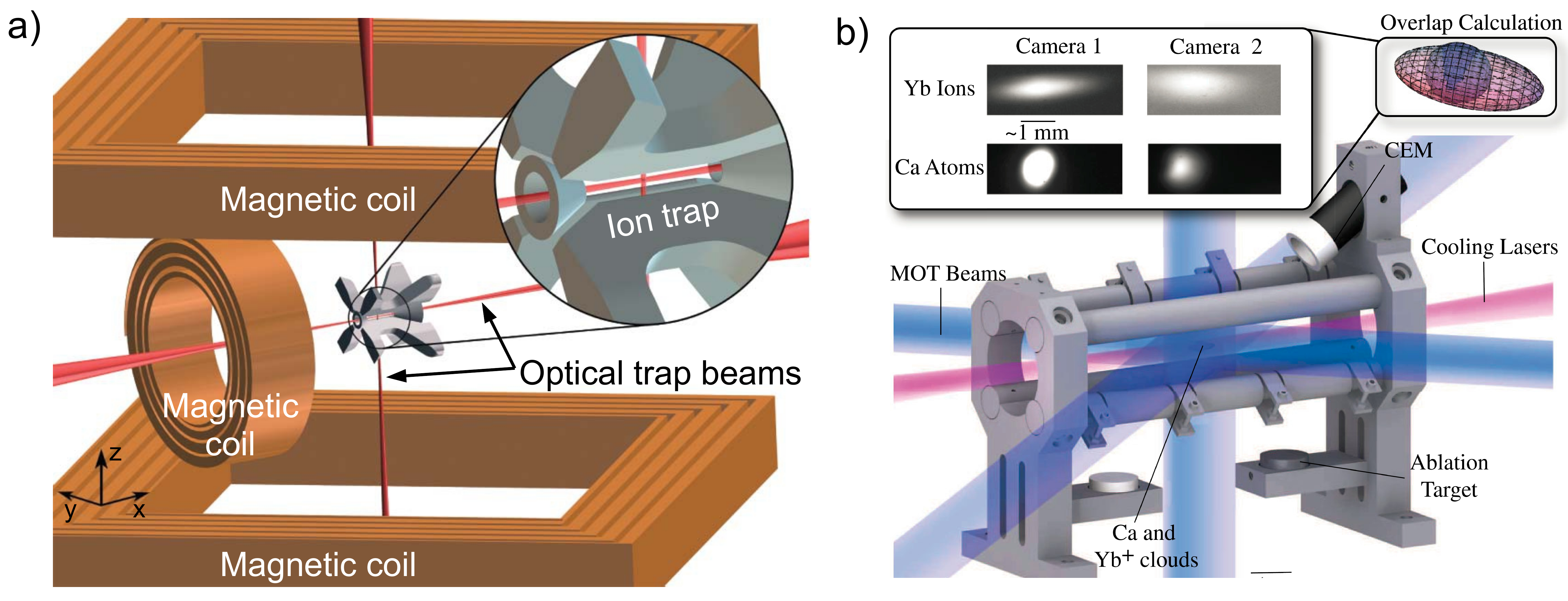,width=\textwidth}}
\caption{Examples of hybrid trap arrangements for cold atoms and ions.   (a) Sketch of the Cambridge setup. The atoms are trapped by using a magnetic trap in the quadrupole-Ioffe configuration
  (QUIC), or an optical dipole trap. The ions are trapped by using a linear Paul trap.   (b) Sketch of the UCLA setup. The atoms are trapped in a magneto-optical trap (MOT), and the ions are trapped in a linear Paul trap.  Figure taken with permission from: W.G. Rellergert, S.T. Sullivan, S. Kotochigova, A. Petrov, K. Chen, S.J. Schowalter, E.R. Hudson, Phys. Rev. Lett. 107, 243201 (2011). Copyright (2011) by the American Physical Society. }
\label{fig1.2SiasKohl}
\end{figure}

The most common way to trap ultracold neutral atoms is by using a Magneto-Optical Trap (MOT) which may overlay with the centre of the ion trap\cite{vuletic09, hudson12_review, willitsch11}
(see Fig. \ref{fig1.2SiasKohl}(b))$.$ This arrangement allows for studying cold collisions between atoms and ions and appeals by its experimental simplicity. In order to perform very precise experiments with a detailed control over the internal and motional quantum states of neutral atoms, this approach has a few shortcomings: First, the experiments always involve resonant light, since in the absence of any further confining potential the atoms remain in their location without laser cooling for at the most a few tens of ms. The resonant light can be opportunely modulated in amplitude in opposition of phase with the ion's cooling light, in order to avoid ionization of the neutral atoms\cite{vuletic09,Hudson11,willitsch11,smith12}$.$ Nevertheless, the state of the atoms and the ions cannot be simultaneously directly controlled during the interactions, but it has to be inferred separately by using for instance from simulations of the optical Bloch equations\cite{willitsch12,hudson12}$.$ Second, the density
of the atomic cloud is not particularly high, on the order of $10^{14}$m$^{-3}$.

Larger densities and full control of the atoms' internal state can be achieved by using a magnetic trap\cite{zipkes11b} or an optical dipole trap\cite{zipkes11b,denschlag12_review} (see Fig. \ref{fig1.2SiasKohl}(a))$.$ An advantage of using a magnetic trap instead of an optical one is that the density of the neutral atoms can be varied typically over a larger range. Small densities ($\lesssim 10^{17}$m$^{-3}$) can be difficult to achieve in a dipole trap since the trap frequency and the trap depth (the maxiumum energy at which a particle remains trapped) can be independently changed only by varying the laser beam waist, i.e., by physically moving an optical element
\footnote{It is possible of course to reduce the density by reducing the number of atoms in the trap. This method in principle allows one to reduce the density to any desirable value. In practice, it is very challenging to measure small atom
clouds, so the uncertainty in the atom density measurement could increase significantly if the atoms number is considerably reduced.}
$.$ On the contrary, in several types of magnetic traps (Ioffe-Pritchard trap, quadrupole-Ioffe configuration (QUIC) trap, Time Orbital Potential (TOP) trap, etc.) frequency and depth are relatively independent parameters. Therefore, these traps offer the largest range of neutral atom densities available, typically from $\lesssim10^{16} $m$^{-3}$ to $10^{20} $m$^{-3}$.

\subsection{Ion trapping}\label{ionsectionSiasKohl}

It is impossible to trap a charged particle in vacuum by using only static electric fields. With an intuitive argument, one can think that in order to create a trapping potential the electric field lines would have to converge to a point in space, violating then the Laplace equation. At the best, one can create a saddle-point potential, which is deconfining in at least one axis. There exist two different designs of ion traps which circumvent this fundamental problem: the Paul - or radiofrequency (RF) - trap\cite{wineland03} and the Penning trap\cite{gabrielse86}$.$

A Paul trap is made of a quadrupole electric potential whose sign is inverted sufficiently frequently, making the charged particle to experience a time-averaged harmonic potential, usually referred as ``secular potential''. This time-average is better valid the shorter the time-scale for the ``flipping'' of the electric potential is. This is made typically smaller then $1\mu$s by using a RF source. Under the action of laser cooling, the ions order in crystal-like structures\cite{wineland87,walther88}$.$ The ions are detected by collecting the fluorescence light they emit, and the photons can be imaged onto cameras or counted by single-photon detectors. The photon detection gives information not only about the number of ions in the trap but it also gives information about some fundamental properties of the particles, like their quantum state (internal and motional) and their temperature\cite{IonReview03}$.$

In a linear design, a Paul trap is made of four cylindrical electrodes connected to the opposite poles of a RF source of frequency $\omega_{RF}$ and amplitude $V_0$. These RF electrodes are parallel and equally spaced, so that they create a quadrupole potential in the plane orthogonal to their symmetry axis
\begin{equation}
\Phi_{RF}=V_0 \frac{x^2-y^2}{2 R_T^2}\sin\left(\omega_{RF} t\right),
\label{RFFieldSiasKohl}
\end{equation}
where $R_T$ is the distance between a RF electrode and the trap centre. The confinement along the direction of the RF electrodes ($z$) is insured by two other electrodes (endcaps) that provide a weak static harmonic potential
\begin{equation}
\Phi_{static}=\frac {m_{ion}}{2 Q} \omega_z^2\left(z^2-\frac{1}{2}\left(x^2+y^2\right)\right),
\end{equation}
where $\omega_z$ is the frequency of the confining potential along the $z$ direction and $Q$ is the ion's charge. The resulting equation of motion is a Mathieu equation, whose solutions can be approximated to the analytic form\cite{wineland03}
\begin{eqnarray}
x(t)&=A_x \sin\left(\omega_x t + \phi_x\right)\left[1+\frac {q}{2}\sin\left(\omega_{RF} t\right)\right],\label{xmotionSiasKohl}\\
y(t)&=A_y \sin\left(\omega_y t + \phi_y\right)\left[1-\frac {q}{2}\sin\left(\omega_{RF} t\right)\right],\label{ymotionSiasKohl}
\end{eqnarray}
where $\omega_{x,y}=\sqrt{\omega_p^2-\frac{1}{2}\omega_z^2}$, $A_{x,y}=\frac{1}{\omega_{x,y}} \sqrt{\frac{2 E_{x,y}}{m_{ion}}}$, $\omega_p=\frac{Q}{\sqrt{2}m_{ion}}\frac{V_0}{R_T^2\omega_{RF}}$, and $q=\sqrt{8}\frac{\omega_p}{\omega_{RF}}$. Here $E_{x,y}$ and $A_{x,y}$ are the energies and the amplitudes of the
ion's motion in the harmonic secular potential, whose frequencies are $\omega_{x,y}$. The analytic solutions eq.(\ref{xmotionSiasKohl}) and eq. (\ref{ymotionSiasKohl}) are valid under the assumption $2 \omega_z^2/\omega_{RF}^2<4 \omega_p^2/\omega_{RF}^2\ll 1$. The typical trap depth is on the order of $1$eV or $10000$K.

Equations (\ref{xmotionSiasKohl}) and (\ref{ymotionSiasKohl}) show that the ion's trajectory has superimposed a high-frequency oscillation at frequency $\omega_{RF}$. This fast motion, called micromotion, is a displacement of the ion proportional to the instantaneous electric field. For this reason, the micromotion amplitude
is non-zero everywhere in space but where the electric field is zero, i.e., ideally at the geometric centre of the trap. However, the presence of spurious DC fields may cause the trapping potential centre and the quadrupole centre not to coincide but to be at a distance $d$.

Keeping $d$ as small as possible is crucial, since if the micromotion amplitude is too large the ion's motion becomes anharmonic and the ion will be eventually lost from the trap. Reducing micromotion can be accomplished by using extra electrodes generating DC electric fields that compensate the effects of spurious fields\cite{wineland98}$.$ This picture is valid in the ideal case, in which the RF field is well described by Eq. (\ref{RFFieldSiasKohl}). In real experiments, however, this description can be affected by a number of practical issues, like a non perfect parallel alignment of the RF-electrodes or phase differences of the applied field between the different electrodes. If this is the case, DC fields alone may not be sufficient to obtain $d=0$, and the micromotion can be corrected only by applying extra AC compensation fields.

Unlike a Paul trap, a Penning trap\cite{gabrielse86} is made of a static electric quadrupole and a static homogeneous magnetic field. When moving out from the quadrupole centre, the ions experience a velocity-dependent Lorentz force that acts as a restoring force. Therefore, the particles remain confined while experiencing closed trajectories in space. In a hybrid system, the collisions with the atoms will alter the ion's trajectory. In a Penning trap, these collisions will affect the amplitude and the direction of the Lorentz force, possibly causing loss of the ions\cite{wineland95}$.$ In a Paul trap, the presence of a real restoring force on the ion makes its use in a hybrid system more promising. Nevertheless, we should ask ourself under what conditions it is possible to trap an ion in a Paul trap in the presence of a buffer gas.

A first answer to this question was given at the early stage of ion trapping\cite{dehmelt68}$.$ Let us consider an ion moving in a Paul trap. The instantaneous velocity can be considered as a sum of the micromotion velocity ${v_{mm}}$ and the secular velocity ${v_{sec}}$. After an elastic collision with an atom at rest, the ion changes its kinetic energy $W$ by the amount
\begin{equation}
\label{collmotionSiasKohl}
\Delta W=m_{at}(1-\cos(\theta))\left(\beta\langle v_{mm}^2 \rangle_{avg}-
\beta^2 \langle v_{sec}^2 + v_{mm}^2 \rangle_{avg}\right),
\end{equation}
where $\beta=m_{ion}/(m_{ion}+m_{at})$, $\theta$ is the scattering angle, and $\langle \cdot \rangle_{avg}$ corresponds to an average over the phase of the RF field. In case $ m_{ion} \ll m_{at}$ the change in energy $\Delta W$ is positive, and the ion heats up. The physical reason for this heating mechanism is that a collision changes the phase between the micromotion and the driving field, whose frequency is typically several order of magnitude larger than the collisional rate. These phase jumps, which are larger for lighter ions, cause some energy coupling between the driving field and the ion's secular motion. As a consequence, the ion heats up, and gets eventually lost from the trap after a number of collisions. This instability of the Paul trap in the presence of a buffer gas depends on the atom-ion mass ratio.

Considering the binary atom-ion collisions as in Eq. (\ref{collmotionSiasKohl}), the unstable regime is expected to begin for\cite{dehmelt68} $m_{ion} \lesssim m_{at}$. More refined analysis\cite{devoe09,zipkes11}$,$ however, give less stringent constraints for a Paul trap to work in the presence of neutral
atoms. This is because the energy distribution of the particles is not a Maxwell-Boltzmann distribution, but rather a power-law distribution\footnote{A power law distribution implies that strictly speaking one can not define a ``temperature'' of an ion. In this chapter, however, we will still refer to ``temperatures'' of an ion when considering the ion's mean energy, after a division by the Boltzmann constant $k_B$}
$,$ whose exponent depends on the atom-ion mass ratio. This distribution may originate from a multiplicative random process, which does not obey the central limit theorem like an additive random process$,$ and where the random events in this case are collisions causing heating of the ion\cite{devoe09}$.$ The tail for large energies of the ion's energy distribution originates then from non-gaussian fluctuations of a series of heating collisions, whose effects are quickly canceled by sympathetic cooling.
Different values of the mass ratio\cite{devoe09} $m_{at} \lesssim 1.55\times m_{ion}$, and\cite{zipkes11} $m_{at} \lesssim 2.17 \times m_{ion}$ were found in two different papers, both confirming that even ions slightly lighter than the atoms can be efficiently trapped in a hybrid system.

\section{Experiments}

It is only recently that experiments have been carried out to investigate atoms and ions jointly at ultracold temperatures. Although the first theoretical proposals date back to more than a decade ago\cite{dalgarno00,smith05}$,$ physicists have been able to implement the hybrid setups only since $2009$\cite{vuletic09}$.$ In these systems the ion is localized in the deep Paul trap, while the atom is part of a large cloud confined in a much weaker potential. Atom-ion interactions lead to distinct physical effects on the components of the hybrid system. On the one hand, the presence of the ion in the atom cloud causes atom losses and temperature increases, due to the different energy scale of the two trapping potentials. On the other hand, the atoms affect the ion in both the internal and the motional state, which can be independently measured.

In this chapter we will show how the effects of atom-ion interactions can be detected by looking at the above mentioned observables. Most of the physical effects that have been observed so far can be well described in terms of binary atom-ion collisions. If the ion's internal state remains unchanged
after a collision with an atom, the collision is called elastic; otherwise it is called inelastic. The main effect of elastic collisions is the sympathetic cooling of an ion in an ultracold gas of atoms\cite{zipkes10}$.$ This ultracold buffer gas cooling is a promising alternative to cool down ion-based quantum computers with respect to laser cooling. Inelastic collisions, however, may lead to the decoherence of a quantum register encoded in ions internal states. The evolution of a single spin-qubit in an ultracold gas has been studied experimentally, and the fundamental decoherence causes have been identified\cite{ratschbacher13}$.$ A special case of inelastic collisions are reaction processes in which a new charged particle with distinct chemical properties is produced, following a charge exchange process\cite{ratschbacher12,Hudson11} or the creation of a molecular ion\cite{willitsch11}$.$ These processes can be controlled in the laboratory by acting on the internal states of the reactants. Finally, in certain cases the effects of atom-ion interactions can be explained only in terms of few-body physics, since binary collisions are insufficient to describe all expected and observed phenomena. The ion can act for instance as a center for few-body reactions like three-body losses of atoms. This phenomenon is detected by observing anomalous atom loss rates\cite{denschlag12}$.$

\subsection{Sympathetic cooling}\label{SympatheticCoolingSiasKohl}

The idea of using a buffer gas of neutral atoms to cool trapped ions goes back to the early days of electric traps for charged particles\cite{langmuir59}$.$ The ion's trapping potential is usually much deeper than the temperature of inert gases injected into a vacuum system, so atom-ion collisions can lead to a continuous cooling of the ion's secular motion\cite{dehmelt68,brincourt83,blatt86}$.$ Collisional cooling makes it possible to cool efficiently even charged
particles whose internal energy levels make laser cooling extremely demanding, like molecules or closed-shell ions\cite{varfalvy01}$.$ In these experiments, the neutral buffer gas is usually a noble gas, which has a closed shell electronic structure making the atoms very stable against inelastic collisions.

Extending neutral buffer gas cooling to ultracold temperatures improves on the techniques used and protrudes into a new physical regime. The temperatures reached with ultracold gases are usually several orders of magnitude lower than the frequency of the ion's potential, making it therefore in principle possible to use a buffer gas to cool an ion down to the ground state of its motion. This regime is interesting for a number of reasons: on the one hand, continuous cooling to the ground state may increase the efficiency of ion-based quantum computers, since the computer operations would have not to be interrupted to cool the ions as it is done in current experiments. On the other hand, the buffer gas can be considered a ``zero temperature'' gas with respect to the ion's energy, and the atom-ion interactions can be studied in a completely new energy range. A number of effects are expected at these energies, including the formation of mesoscopic molecular ions\cite{cote02} and the presence of atom-ion Feshbach resonances \cite{idziaszek09}$.$

\begin{figure}
\centerline{\epsfig{figure=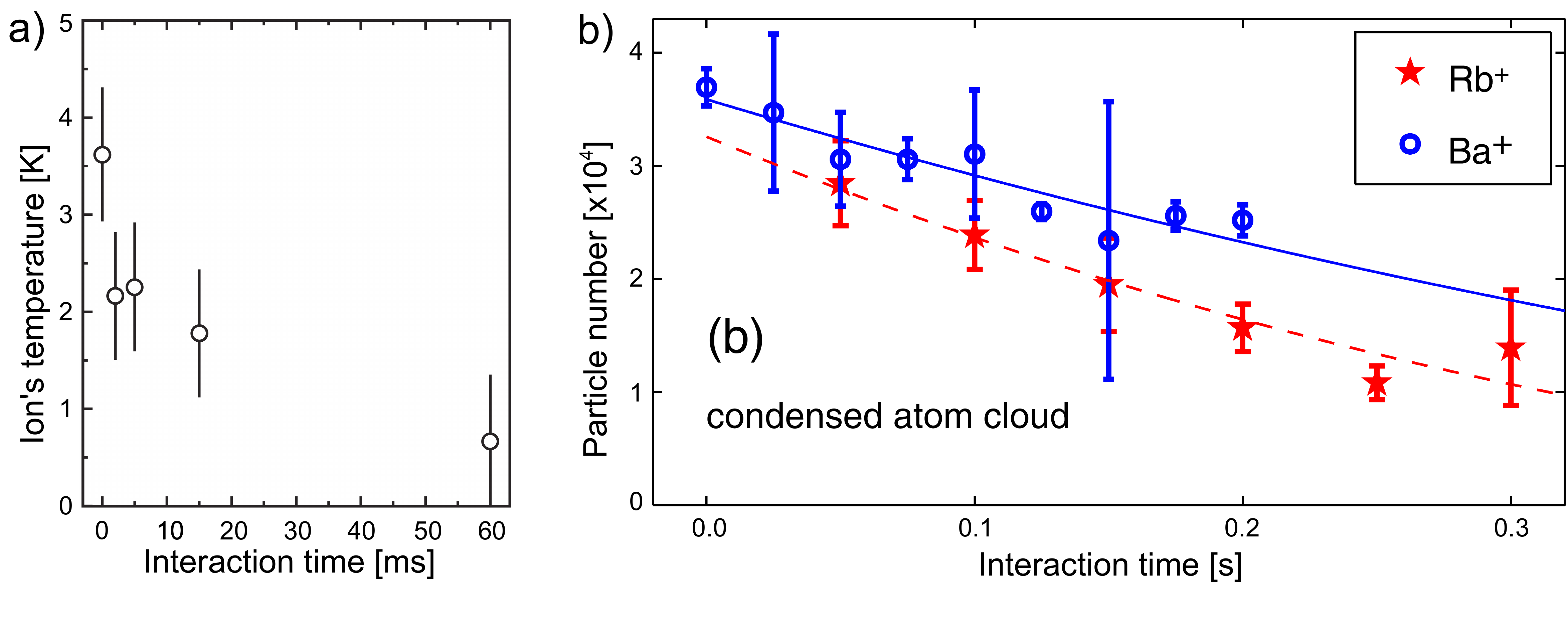,width=\textwidth}}
\caption{ a) Sympathetic cooling of a single trapped ion in an ultracold atomic gas. A ``hot'' Yb$^+$ ion is immersed in a Bose-Einstein condensate
of Rb atoms, and the ion's temperature for different interaction times is measured monitoring the ion's fluorescence. The lowest value of the ion's temperature is comparable to the minimum temperature measurable with this method. Figure taken from ref. \cite{zipkes10}$.$ b) Depletion of a Bose-Einstein condensate of Rb atoms due to atom-ion collisions. The depletion was observed using two different ions: Ba$^+$ and Rb$^+$. Figure taken with permission from: S. Schmid, A. H{\"a}rter, J. Hecker Denschlag, Phys. Rev. Lett. 105, 133202 (2010). Copyright (2010) by the American Physical Society.}
\label{figelasticSiasKohl}
\end{figure}

Buffer gas cooling by use of an ultracold gas was first demonstrated using a single trapped ion immersed into a Bose-Einstein condensate\cite{zipkes10}$.$ A single $^{174}$Yb$^+$ ion at $T=7K$ was immersed into a Bose-Einstein condensate of $\sim5\times10^4$ atoms of $^{87}$Rb at $T\sim100$nK, and the temperature of the ion was measured at different times after the immersion by using the Doppler fluorescence technique\cite{wesenberg07} (see Fig. \ref{figelasticSiasKohl}(a)). The ion's temperature was reduced in a few tens of milliseconds to the measurement resolution limit of $\sim 500$mK. A cooling efficiency of almost 1000 vibrational quanta per collision was observed.

Ion cooling is accompanied by a depletion of the Bose-Einstein condensate\cite{zipkes10,denschlag10} (see Fig. \ref{figelasticSiasKohl}(b)). The atom loss is a consequence of the difference in the energy scale of the ion and the atoms potentials. This difference is typically such that the frequency of the ion trap is of the same order of magnitude of the depth of the atoms' trap ($\sim 1 \mu$K). In experiments with deeper atoms trap, an increase of the atoms' temperature is also observed, mainly induced by forward scattering\cite{zipkes10b}$.$

Under which conditions and down to what temperature is it possible to cool an ion with buffer gas cooling? It seems in principle possible to cool an ion down
to the absolute ground state of the secular potential, since the temperature of the buffer gas is typically lower than the ground state energy of the ion's potential. Nevertheless, as shown in section \ref{ionsectionSiasKohl}, in the presence of micromotion it is possible to couple energy from the RF electric field to the secular motion through collisions with the atoms. Therefore, the ion's mean energy will result from an ``equilibrium'' between the sympathetic cooling and the micromotion-induced heating rates. The energy scale of the ion temperature can then be considered the micromotion energy:
\begin{equation}
\label{micromotionSiasKohl}
E_{mm}=\frac{1}{2}\omega_{RF}^2 a_{mm}^2,
\end{equation}
where $a_{mm}$ is the micromotion amplitude. The dependence of the ion's temperature on the micromotion amplitude has been verified in several experiments\cite{zipkes10b,denschlag13}$.$ In these experiments, the micromotion was varied by applying a radial offset field $\mathcal{E}_{r}$ in order to obtain a micromotion amplitude $a_{mm}=\sqrt{2}q {\mathcal{E}_{r}} /{\left( m_{ion} \omega_{sec}^\perp \omega_{RF}\right)}$, where $\omega_{sec}^\perp$ is the radial frequency of the secular motion. The effects of the micromotion were observed both by looking at the ion's temperature$,$ and by looking at the atoms temperature increase and loss rate in the presence of the ion. A linear increase of the ion's mean energy with respect to $\mathcal{E}_{r}^2$ has been observed\cite{zipkes10b}$.$ By measuring the effects of the ion on the neutral atoms, a compensation of the stray DC fields up to $0.02$V/m has been reported\cite{denschlag13}$.$ The corresponding energy, however, is currently still far from ground state cooling.

Nevertheless, there is no physical limitation preventing from a better compensation of micromotion and therefore from reaching lower temperatures. By using an adequate number of electrodes in the trap, one could in principle compensate for all possible contributions to micromotion. An ultracold regime for atom-ion physics would be extremely interesting to reach, since there are several intriguing questions still open over the possibility of reaching the ion's ground state with collisional cooling. One effect relates to the long-range attractive interaction between atoms and ions, which extends far beyond the typical length scale of the wave function of the trapped ion in its vibrational ground state. Owing to the long-range interaction, the atom could pulls the ion out of the trap centre upon approach \cite{vuletic12}. This could induce a non-adiabaticity of the motion of the ion with respect to the RF field. The tuning parameter for the strength of this effect is the mass imbalance as well as the trap stiffness. Another distinct point of novelty in low-energy atom-ion collisions is that at sufficiently low temperatures the de-Broglie wavelength of the ion becomes larger than the micromotion amplitude, and a kind of Lamb-Dicke regime for the micromotion would be reached. In this regime the micromotion energy would couple to the atom and the ion after a collision only in units of $\omega_{RF}$, since the continuous energy spectrum of the micromotion-coupled energy would be rather constituted by energy sidebands. 

There exist several strategies to reach lower temperatures in atom-ion systems. A first alternative cooling method is the swap cooling\cite{rangwala12}$.$ In this mechanism, a hot trapped ion and a cold neutral atom undergo an inelastic collision in which one unit of charge is exchanged between the colliding particles. A charge exchange process gives a net gain in the temperature of the charged particle, since before the collision the atom is typically orders of magnitude colder than the original ion. This swap cooling mechanism is strongest when the ion and the atoms are of the same isotope, since the charge exchange rate is equal to $0.5$. Other alternatives to attain lower ion's temperature are the use of different ion traps like a micromotion-reduced 22-poles trap\cite{gerlich95}$,$ or a micromotion-free purely optical trap \cite{schaetz10}$.$

Finally, cold collisions in hybrid atom-ion systems can be used not only to sympathetically cool the ion's motional energy, but even to reduce the vibrational temperature of molecular ions. This sympathetic vibrational cooling has been observed with BaCl$^+$ molecules immersed in a cold cloud of Ca atoms\cite{hudson13}$.$ This experiment is a very promising alternative with respect to internal-state laser cooling of molecules, and opens the way to using hybrid atom-ion systems in molecular and chemical physics.

\subsection{Atom-Ion chemical reactions}

In this section we will consider the effects of inelastic collisions in which the ion undergoes a chemical reaction or changes its internal state. Due to the cold temperatures of both ion and atoms in a hybrid quantum system, we will consider exothermic processes only. In binary collisions, the inelastic phenomena that may be observed are of three kinds:
\begin{enumerate}[(a)]
\item Charge exchange reactions: $X^+ + Y \rightarrow X + Y^+$ \label{ceSiasKohl}
\item Creation of molecular compounds: $X^+ + Y \rightarrow XY^+$ \label{moleculeSiasKohl}
\item Collisional quenching: $\left( X^+ \right) ^* + Y \rightarrow X^+ + Y$
\end{enumerate}
where the notation $\left(\cdot\right)^*$ indicates a particle in an excited electronic state.

Conservation of energy implies that the energy difference between the input and the output channels has to be released by emission of a photon or by the motional heating of the colliding particles. Non-radiative processes may lead to ion loss, while in radiative processes the small recoil energy of the photon insures that a charged particle remains kept in the ion trap. Non-radiative inelastic collisions can occur when two different molecular potentials are coupled at a certain interparticle separation, creating an avoided crossing. If this is the case, the colliding particles' input and output channels may differ if a quantum tunneling event occurs at the avoided crossing. Radiative processes can happen if the Franck-Condon overlap between molecular wave functions is sufficiently large. In both cases the particles need, in first approximation, to approach each other to a sufficiently short distance. For this reason inelastic processes occur at a rate proportional to the Langevin rate.

\begin{figure}
\centerline{\epsfig{figure=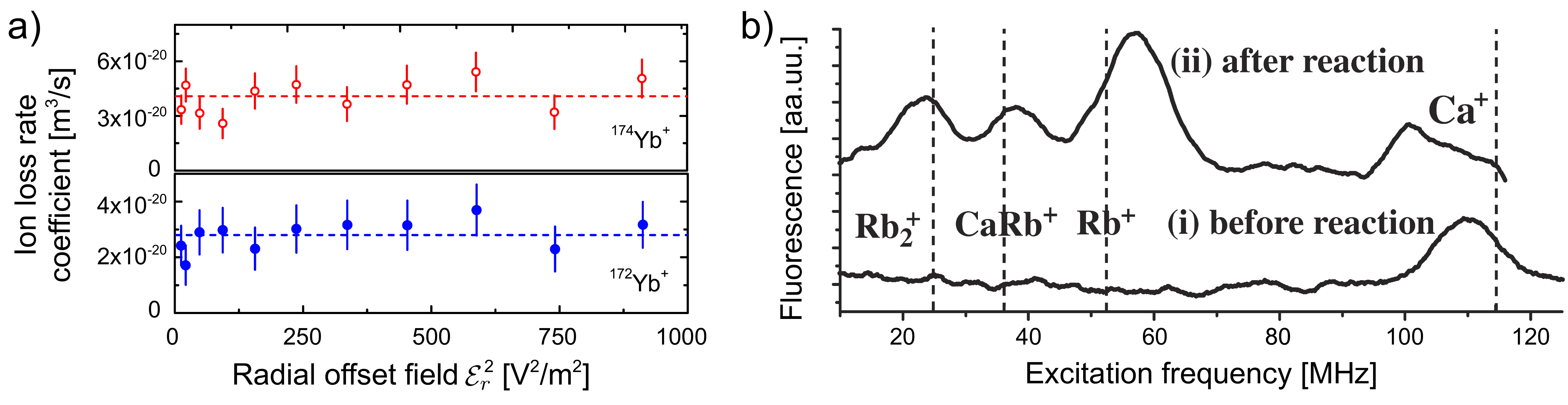,width=\textwidth}}
\caption{Non-resonant charge exchange rate vs collision energy. The mean ion energy was changed by applying a radial offset field $\mathcal{E}_{r}$. The charge exchange signature was the loss of the ion's fluorescence signal. The ion loss rate was found to be independent on the mean ion energy, as expected for Langevin collision rate. The data were taken for two different isotopes of the Yb$^+$ ion. Figure taken with permission from: C. Zipkes, S. Palzer, L. Ratschbacher, C. Sias, M. K{\"o}hl, Phys. Rev. Lett. 105, 133201 (2010). Copyright (2010) by the American Physical Society. b) Mass spectrometry of a large Ca$^+$ ion crystal before and after reacting with a cold cloud of Rb atoms. The spectrometry is performed by applying an external field at a variable frequency to the trap electrodes, and by looking at the modification of the Ca$^+$ fluorescence signal. The spectrum taken after the interaction shows three peaks corresponding to the excitation frequencies of Rb$_2^+$, CaRb$^+$ and Rb$^+$ ions. The presence of CaRb$^+$ in the crystal is the direct evidence of molecule formation due to atom-ion collisions. Figure taken with permission from: F.H.J. Hall, M. Aymar, N. Bouloufa-Maafa, O. Dulieu, S. Willitsch, Phys. Rev. Lett. 107, 243202 (2011). Copyright (2011) by the American Physical Society.}
 \label{chargeexchangeSiasKohl}%
\end{figure}

The first evidence of an energy independent inelastic rate in the regime of cold collisions was obtained\cite{vuletic09} for quasi-resonant charge exchange collisions between different isotopes of Yb and Yb$^+$. In this experiment, up to a few thousand ions were immersed in a cold cloud of neutral atoms, and the charge-exchange processes were detected by observing a reduction of the fluorescence light emitted by the ions. In the non-resonant case, a measurement of the energy dependence of the inelastic collision rate was performed with single ions by using Rb atoms and two different isotopes of Yb$^+$ ions \cite{zipkes10b} (see Fig. \ref{chargeexchangeSiasKohl}(a)). In this experiment, the atoms' and the ion's internal states were precisely determined, since the collisions were occurring in absence of resonant light. After letting the atoms and the ion interact, the atoms were removed from the trap, and the ion was illuminated with resonant light. The absence of fluorescence light was the signature of the occurrence of an inelastic collision after which the ion was either lost from the trap or a different charged particle was trapped. For atoms and ions in the lowest electronic energy level, the inelastic collisional rate was found to be five orders of magnitude lower than in the quasi-resonant charge exchange case.

More information about the inelastic collision outcomes can be obtained by using an ion crystal instead of a single ion. Let us consider the simplest case of a crystal made of two ions, in which one of the two ions undergoes an inelastic process while the second one remains in its original state. This unchanged ion will provide some fluorescence light under resonant light illumination. By detecting the fluorescence light with a CCD camera, it is possible to observe the spatial position of the unchanged ion. A dark ion in the trap can therefore be indirectly detected since the Coulomb force will displace the unchanged ion from the trap centre. In case a dark ion is detected, its mass can be measured in a spectroscopic measurement by modulating the trapping potential\cite{Schiller07} or the cooling light intensity\cite{Drewsen04} at a frequency $f$. When the excitation frequency $f$ matches a collective vibration mode of the Coulomb crystal, the crystal heats up causing a Doppler shift of the fluorescence. Depending on the detuning of the cooling light, the heating of the ion crystal will result in a decrease\cite{softley08} or an increase\cite{Schiller07} of the fluorescence light detected.

An example of a mass spectrometry analysis of the reaction products in a Rb+Ca$^+$ system\cite{willitsch11}  is shown in Fig. \ref{chargeexchangeSiasKohl}(b). A peak at the mass of CaRb$^+$ molecule was observed after Rb and Ca$^+$ were made interact. The presence of a peak corresponding to Rb$^+_2$ molecules indicates the occurrence of secondary collisions of the kind CaRb$^+$ $+$ Rb $\rightarrow$ Rb$^+_2$ $+$ Ca, while the peak corresponding to Rb$^+$ mass attests the charge-exchange reactions between Ca$^+$ and Rb, leading to Rb$^+$ ions.

Hybrid systems of atoms and ions enable the observation of chemical reactions like charge-exchange and molecule formations at the single particle level. But how much control can be exerted on atom-ion chemical reactions? A first ``knob'' in the hands of an experimentalist is the initial choice of the elements used in the experiment. So far, a number of combinations of elements have been studied, including: Rb+Yb$^+$, Yb+Yb$^+$, Rb+Rb$^+$, Rb+Ba$^+$, Ca+Yb$^+$, Rb+Ca$^+$, Rb+N$_2^+$, Ca+Ba$^+$, Na+Na$^+$, Ca+BaCl$^+$. Once the combination is set, a second knob is the control over the internal state of the colliding particles. In fact, particles in different internal states experience different molecular potentials during the collision, causing a variation of the chemical reaction rate and of the reaction outputs. The dependence on the internal state in cold atom-ion chemical reactions has been observed in several atom-ion combinations, including Rb+Yb$^+$, Rb+Ca$^+$, Rb+N$_2^+$, and Ca+Ba$^+$.

The use of this dependence to control chemical reactions was demonstrated first for a Rb+Yb$^+$ system\cite{ratschbacher12}$.$ The Yb$^+$ ion offers a number of excited electronic energy levels whose lifetime is much longer than the inverse of the Langevin collision rate, while Rb atoms offer two hyperfine levels split by $\sim 6.8$ GHz. The internal state of the atoms and the ion was then precisely determined in the collisions by using laser and microwave sources. A variation of the inelastic collision rate over five orders of magnitude was observed. In particular, when optically pumped in the $^2D_{3/2}$ state, the Yb$^+$ ion undergoes a chemical reaction for basically every Langevin collision. This very high reactivity could be used to perform quick, local measurements of the atomic density\cite{kohl07}$,$ since the reaction was shown to depend linearly on the atom density over more than two orders of magnitude. Moreover, both the chemical reaction rate and the branching ratio between radiative/non-radiative processes showed a dependence on the atoms' hyperfine state. Similar results were obtained in experiments using\cite{willitsch12} Rb+N$_2^+$ and\cite{hudson12} Ca+Ba$^+$. In the latter case, though, no dependence on the atom internal state was observed.

\subsection{Few-body physics in atom-ion hybrid systems}

In the experiments described in the previous sections, only binary collisions were taken into account. However, atoms and ions at cold temperatures also exhibit three- or many-body effects since the atom-ion interaction potential is long-ranged and attractive.

Three-body physical effects in an atom-ion system have been observed in a Rb-Rb$^+$ experiment\cite{denschlag12}$.$ In this experiment the micromotion was compensated well enough to lower the energy of the ion below $1$mK. Under these conditions, the Rb$^+$ ion catalyzed the formation of Rb$_2$ molecules that were created by the association of two Rb atoms. In these three-body processes the role of the ion is twofold. On the one hand the ion causes an increase of the local atom density through the attractive, long-ranged interaction potential. On the other hand, the ion plays the role of a ``third body'' that enables the conservation of both energy and momentum in the Rb$_2$ molecule formation. The signature of this effect in the experiment was the heating of the ion at energies up to several $0.1$eV. This heating was detected by observing a sudden decrease in the atom loss rate, caused by the fact that the ``hot'' ion moved to a trajectory much larger than the atomic cloud.

While three-body effects have been succesfully observed, many body effects of atoms and ions in hybrid system are still elusive. In the case of ultracold collisions, the formation of mesoscopic molecular ions is suggested\cite{cote02}$.$ The atoms are predicted to fall into the lowest bound molecular state of the atom-ion potential, dissipating the excess energy in a phonon emitted in the ultracold gas. This effect has never been observed, though, since the temperatures that are currently available are still too large to access s-wave collisions.

\subsection{Quantum coherence}

Trapped ions are an extremely important resource for precision measurements and quantum computation. Both the most precise clock currently available and the quantum computer with the largest number of qubits to-date are based on trapped ions. These achievements were possible thanks to the possibility of manipulating the quantum state of an ion. A single trapped ion can realize a spin-1/2 system, i.e. the most fundamental quantum mechanical object.

A single spin-1/2 system interacting with an environment is a crucial paradigm to describe how elementary quantum mechanical objects decohere. This is of both conceptual interest, since it links quantum mechanics to the classical world, and of practical interest because it limits applications of quantum mechanics, such as quantum computers and atomic clocks. The spin-dynamics and the decoherence arising when a spin-1/2 interacts with an environment
determine its potential use as a qubit and are responsible for a multitude of impurity effects encountered in the solid state. While an extensive amount of theoretical work on this problem exists\cite{Leggett87,prokofev00}$,$ experiments with well-controlled and adjustable environments are scarce.

A hybrid quantum system of atoms and ions is an ideal ground to explore the coherent evolution of a single spin-qubit in a bath, since the ion-based spin-qubit can evolve in an atom-based environment of tunable spin states and tunable density. The first study of the coherent evolution of a single trapped ion in an ultracold bath of atoms was realized in a Rb-Yb$^+$ system\cite{ratschbacher13}$.$ A spin-qubit state $|\psi\rangle_i= \alpha |\uparrow\rangle_i + \beta |\downarrow\rangle_i $, where the subscript $i$ refers to the ion, $|\alpha|^2+|\beta|^2=1$, and $\alpha,\beta\in \mathbb{C}$, was encoded in a Yb$^+$ ion using two alternative basis. In a first method, the magnetic insensitive hyperfine levels $|F=0, m_F=0\rangle_i$ and $|F=1, m_F=0\rangle_i$ of the electronic ground state of the $^{171}$Yb$^+$ ion were used. A second strategy was to realize a ``Zeeman-qubit'' by using the two sublevels $|J=1/2,m_J=1/2\rangle_i$ and $|J=1/2,m_J=-1/2\rangle_i$ created in a $^{174}$Yb$^+$ ion after applying a magnetic field.

\begin{figure}
\centerline{\epsfig{figure=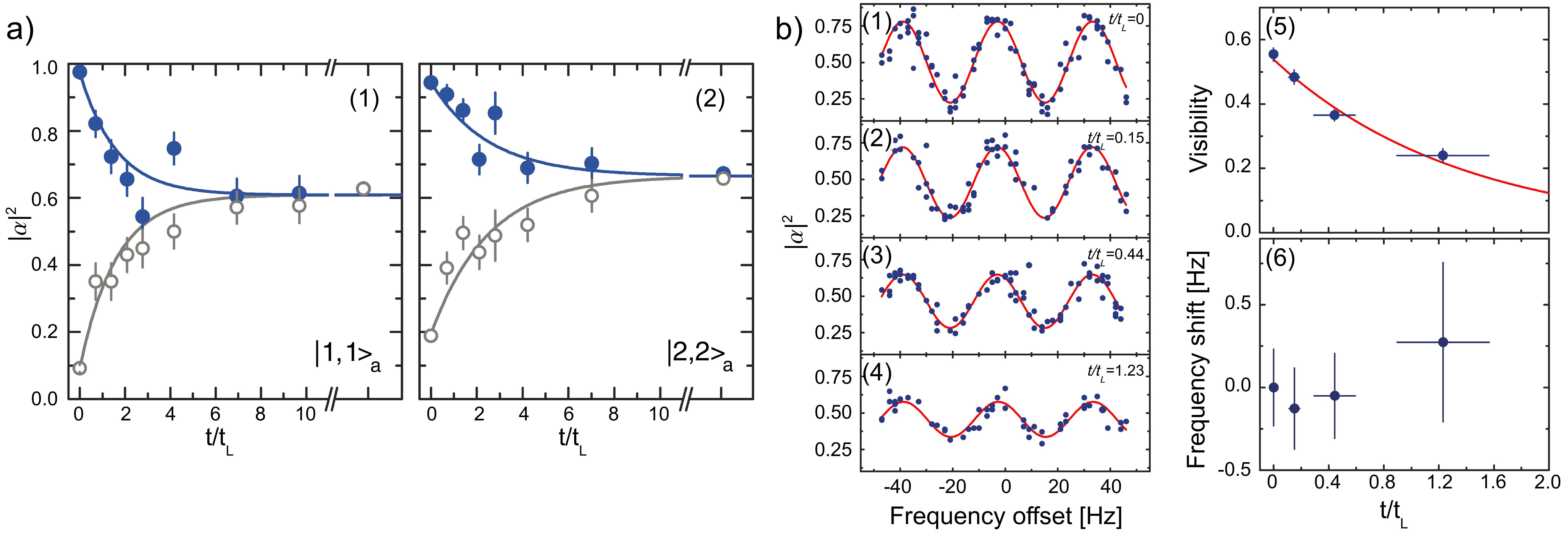,width=\textwidth}}
\caption{Decoherence of a single-ion qubit in an ultracold neutral atom cloud. a) Measurement of the probability $|\alpha|^2$ of finding the ion in the $|\uparrow\rangle_i$ state at different times $t$ after immersion in an ultracold gas of Rb atoms. The filled (empty) circles correspond to an ion initially prepared in the $|\uparrow\rangle_i$ ($|\downarrow\rangle_i$) state. The atom population decreases to a steady state that depends on the atoms internal state. Measurements are shown for two different internal states of the neutral atoms (1) and (2). b) Coherent evolution of a single spin qubit in an environment of ultracold atoms. (1)-(4) Ramsey fringes taken at different values of $t/t_L$. Here the interaction time was fixed, and the Langevin time $t_L$ was changed by varying the atomic density. (5) Ramsey fringes visibility as a function of $t/t_L$. The effect of decoherence causes the visibility to drop in a time comparable to the Langevin collision time $t_L$. (6) Measurement of the resonant qubit clock frequency at different atomic densities. No frequency shift was observed within the experimental errors. Figure taken with permission from: L. Ratschbacher, C. Sias, L. Carcagn{\`i}, J. Silver, C. Zipkes, M. K{\"o}hl, Phys. Rev. Lett. 110, 160402 (2013). Copyright (2013) by the American Physical Society.}
 \label{coherenceSiasKohl}%
\end{figure}

In a first measurement (see fig. \ref{coherenceSiasKohl}(a)), the ion was prepared in a $|\uparrow\rangle_i$ or $|\downarrow\rangle_i$ state, and then it was inserted into the ultracold atom cloud. The initially polarized ion spin relaxed into a mixed steady-state with a characteristic time $T_1$ equal to a few Langevin collision times. By measuring the ion's spin steady state for different internal states of the atomic bath, it was possible to identify two distinct mechanisms causing the decoherence of the ion's spin. The first process are spin-exchange collisions, in which a unit of angular momentum is exchanged between an atom and ion in a collision. The second process are spin-relaxing collisions, in which the spin and the orbital motion couple, causing a transfer of angular momentum from the motional state to the spin state. This spin-nonconserving mechanism was responsible for the decoherence of the ion's state $|\uparrow\rangle_i$ in an environment of atoms polarized in the stretched state $|F=2,m_F= 2\rangle_a$, which otherwise should not have occurred (here the subscript $a$ refers to an atom's state).

In a second measurement, the ion was prepared in a superposition state $\left(  |\uparrow\rangle_i + i |\downarrow\rangle_i \right) / \sqrt{2} $, and its coherence time was measured by looking at the visibility of the Ramsey fringes. The characteristic time $T_2$, i.e. the decay constant of the off-diagonal terms of the ion's density matrix, was measured to be of the same time scale of the characteristic time $T_1$ (see Fig. \ref{coherenceSiasKohl}(b)). In this measurement, the spin qubit was encoded in the hyperfine levels of the $^{171}$Yb$^+$ ion. Interestingly, when the atomic bath was prepared in the more energetic $|F=2,m_F=2\rangle_a$ state, the spin-nonconserving spin-relaxing collisions were observed even for this spin-qubit, whose energy levels are separated by $12.6$GHz. This energy, which is one order of magnitude larger than the micromotion energy of the ion, was partially provided by the atoms' internal energy of $6.8$GHz. This mechanism of distribution of the atoms' internal energy to the kinetic and then internal energy of the ion was proved by performing the same experiment with the atoms in the less energetic $|F=1,m_F=-1\rangle_a$ state. In this latter case, no spin-relaxation of the lowest $|F=0,m_F=0\rangle_i$ state was observed. Finally, the frequency of the qubit clock transition in $^{171}$Yb$^+$ showed no variation due to the presence of the ultracold bath. The measurement resolution set by the spin relaxation rate was $\Delta \nu=4 \times 10^{-11} E_{i}^{(HF)}/h$, where $h$ is the Planck's constant, and $E_{i}^{(HF)}\simeq h \times 12.6$ GHz is the energy difference of the ion's hyperfine clock transition $|F=0, m_F=0\rangle_i \rightarrow |F=1, m_F=0\rangle_i$

\section{Outlook}

Experimental research on atom-ion hybrid systems is a novel field with a great space to explore ahead of us. The hybrid system crosses the borders between atomic physics, chemistry and quantum information processing and promises to gain fundamental understanding of cold chemistry as well as system-bath interactions in many-body physics. Already now, a number of experiments have considerably improved our understanding of the physics of collisions between atoms and ions at low temperature. In the future, a better control of the atom-ion interaction would be desirable, which is anticipated by means of Feshbach resonances\cite{idziaszek09}, which have yet escaped experimental observation. Lower collisional temperatures would disclose the physics of s-wave collisions and of atom-ion interactions beyond the semiclassical approximation. It might then be possible to observe ground state buffer gas cooling and the formation of mesoscopic molecular ions\cite{cote02}$.$ At this higher level of control, the ion could be used as an atom probe in a cloud loaded in a three dimensional optical lattice\cite{kohl07}$,$ acting as a ``microscope tip'' in a many-body state and in a quantum register.

\section{Acknowledgements}

We are indebted to Stefan Palzer, Christoph Zipkes, Lothar Ratschbacher, Leonardo Carcagn{\`i} and Jonathan Silver for their work in designing, building, and operating the Cambridge hybrid atom-ion experiment, and for the countless discussions of physics shared over several years.
This work has been supported by the Alexander-von-Humboldt Professorship, EPSRC (EP/H$005676/1$), ERC (Grant No. $240335$), the Leverhulme Trust (C.S.), the Royal Society, and the Wolfson Foundation.

\bibliographystyle{ws-rv-van}
\bibliography{chapterKohl}

\end{document}